\documentclass[aip, amsmath, amssymb, reprint]{revtex4-1}

\usepackage{graphicx}
\usepackage{dcolumn}
\usepackage{bm}

\begin{document}

\title{Gain assisted nanocomposite multilayers with near zero permittivity modulus at visible frequencies}

\author{Carlo Rizza}
\affiliation{Dipartimento di Fisica e Matematica, Universit\'a dell'Insubria, Via Valleggio 11, 22100 Como, Italy} \affiliation{Consiglio Nazionale
delle Ricerche, CNR-SPIN, 67100 Coppito L'Aquila, Italy}

\author{Andrea Di Falco}
\affiliation{School of Physics and Astronomy, University of St. Andrews, North Haugh, St. Andrews KY16 9SS, UK}

\author{Alessandro Ciattoni}
\email{alessandro.ciattoni@aquila.infn.it}
\affiliation{Consiglio Nazionale delle Ricerche, CNR-SPIN, 67100 Coppito L'Aquila, Italy}

\date{\today}

\begin{abstract}
We have fabricated a layered nano-composite by alternating metal and gain medium layers, the gain dielectric consisting of a polymer incorporating
optically pumped dye molecules. Exploiting an improved version of the effective medium theory, we have chosen the layers thicknesses for achieving a
very small value of the real part of the permittivity $\epsilon_\|$ (parallel to the layers plane) at a prescribed visible wavelength. From standard
reflection-transmission experiments on the optically pumped sample we show that, at a visible wavelength, both the real and the imaginary parts of
the permittivity $\epsilon_\|$ attain very small values and we measure $| \epsilon_\| | = 0.04$ at $\lambda = 604$ nm, amounting to a $21.5$-percent
decrease of the minimum $| \epsilon_\| |$ in the absence of optical pumping. Our investigation thus proves that a medium with a dielectric
permittivity with very small modulus, a key condition which should provide efficient subwavelength optical steering, can be actually synthesized.
\end{abstract}


\maketitle

Epsilon-near-zero (ENZ) metamaterials are engineered materials with a very small real part of the dielectric permittivity. Their unique properties,
enabling exotic light behavior, have recently been addressed both in the linear \cite{Silver,AluAlu,Nguyen} and in the nonlinear regime
\cite{FengFe,Ciatt1,Ciatt2,RizzaR}. The ENZ condition can be easily achieved by alternating materials with negative (metals) and positive
(dielectric) values of the real permittivity, to obtain an average value close to zero, for a given wavelength and polarization \cite{KarriK,Robert}.
This effective medium approximation is only valid if the individual layers are much smaller than the wavelength, i.e. down to few nanometers in the
optical range. However, for practical applications the use of a few nanometers thin metal layers has to be generally avoided, since size-effects are
responsible for substantial material absorption, much greater than that characterizing the metal bulk. As a result, it is necessary to choose a
sub-optimal geometrical configuration, and adopt a more refined theoretical model able to describe it, accounting for nonlocal corrections
\cite{ElserE,Pollar}.

On the other hand, while most of the research effort on ENZ metamaterials is focused on achieving a very small real part of the permittivity, we
point out that its imaginary part plays a crucial role as well. Hence, the quantity to be minimised is $|\epsilon|$. In such generalized ENZ
metamaterials, small permittivity changes would no longer be playing the role of a mere perturbation and can have a dramatic impact on optical
propagation \cite{Ciatt1}. This novel condition would allow an extremely efficient external control (e.g. via optical nonlinearity, electro-optic
effect and acusto-optic effect), for example to obtain very effective (low power) optical steering and light manipulation \cite{Ciatt3}. It is then
clear that to exploit the full potential of generalized ENZ metamaterials, managing the imaginary part of the permittivity, i.e. adopting schemes for
loss compensation is of pivotal importance. Various techniques and gain mechanisms have been proposed so far to achieve this goal. Examples are
negative index materials with gain nanoconstituents \cite{KlarKl,Wuestn,XiaoXi}, negative dielectrics hosting quantum dots
\cite{Bratko,DongDo,FuFuFu,Holmst,WebbWe} and surface plasmon polaritons excitation in the presence of optically pumped dye molecules
\cite{DeLeo1,DeLeo2,Nogin1,Nogin2}.

\begin{figure}
\includegraphics[width=0.45\textwidth]{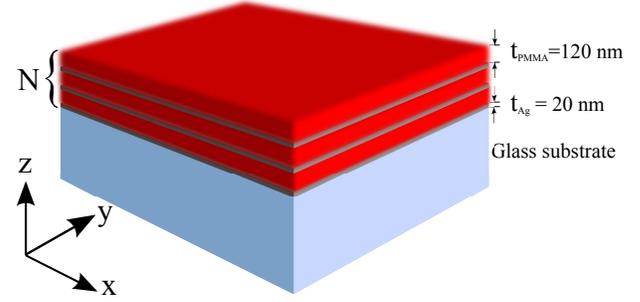}
\caption{(Color online) Geometry of the samples. The layered composite lies on a glass substrate and it is made up of N bilayers, each one comprising
a silver and a PMMA layer of thicknesses $t_{Ag}$ and $t_{PMMA}$, respectively}
\end{figure}

In this letter we demonstrate experimentally that a metallo-dielectric nano-laminate can be designed to fulfil the generalized ENZ condition, for the
polarization parallel to the layers plane ($|\epsilon_\| | \ll 1$), at visible frequencies. We achieve control over $|\textrm{Re}(\epsilon_\|)|$ by
carefully choosing the layers thickness, whereas we reduce $|\textrm{Im}(\epsilon_\|)|$ by using an optically pumped organic dye. Most remarkably,
our experimental findings are perfectly predicted by the improved version of the effective medium theory, after Elser $\textit{et al.}$ \cite{ElserE}

The multilayer stacks were fabricated alternating layers of e-beam evaporated Silver ($20$ nm) and spun polymethilmethacrylate (PMMA 495, from
Microchem) ($120$ nm), on glass. The layers thickness was measured with a Dektak stylus profiler, with accuracy better than $1$ nm. The polymer
included an optimized concentration of 1.5 mg/ml of Lumogen RED dye, with absorption peak at $565$ nm and emission peak at $\sim 610$ nm. Figure 1 shows
a sketch of a typical multilayer sample. The optimization was performed with independent experiments, by measuring with an integrating sphere the
emission efficiency of samples with a $100$ nm thick film, with different concentrations, from $1$ mg/ml to $6$ mg/ml. We fabricated several samples
with varying number of metal/polymer bilayers, from 3 to 5. The transmission (T) and reflection (R) measurements were performed with polarized
broadband light emitted by a tungsten lamp, all with normal incidence to the samples. T and R were collected with two independent silicon-based optical
spectrometers, limiting the observed wavelength range to $500-1100$ nm.  The dye was then excited with a doubled Nd:Yag emitting cw light at $532$ nm,
with an average power of $200$ mW. The samples were pumped at an angle ($\sim 40^o$ with respect to the normal) to avoid direct pump light reaching either of the
spectrometers. To prevent the tungsten lamp from exciting the dye, we filtered the incoming probe signal with a long pass filter, with cutoff
wavelength at $570$ nm.

\begin{figure}
\includegraphics[width=0.45\textwidth]{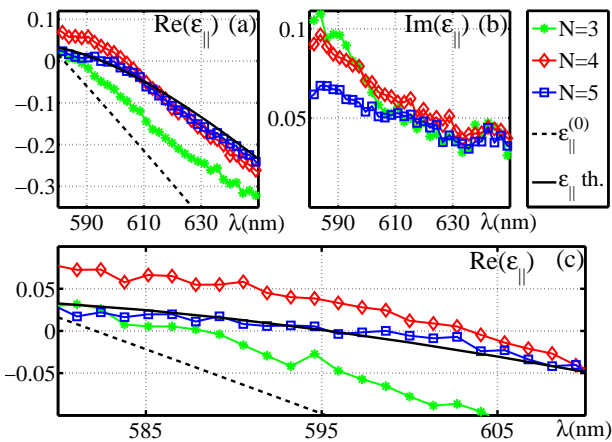}
\caption{(Color online) Retrieved dielectric permittivity $\epsilon_\|$ of the samples, without optical pumping, made up of $N=3,4,5$
metallo-dielectric bilayers (solid lines with markers) as a function of the vacuum wavelength $\lambda$. In panel (a) $\textrm{Re}(\epsilon_\|)$ is
plotted together with theoretical $\textrm{Re}(\epsilon_\|)$ evaluated from Eq.(\ref{permitt}) (solid lines without markers) and
$\textrm{Re}(\epsilon_\|^{(0)})$ obtained from the standard effective medium theory (dashed line). In panel (b) $\textrm{Im}(\epsilon_\|)$ is
reported whereas panel (c) is a magnified copy of panel (a) in the spectral range $580$ nm $ < \lambda < 610$ nm within which
$\textrm{Re}(\epsilon_\|)$ changes sign.}
\end{figure}

To design our samples, we theoretically modelled the medium dielectric tensor
$\epsilon=\textrm{diag} \left( \epsilon_\|,\epsilon_\|,\epsilon_\perp \right)$, where according to ref. [\onlinecite{ElserE}]

\begin{equation} \label{permitt}
\epsilon_\|= \frac{\epsilon_\|^{(0)}}{\displaystyle  1 - \frac{(\epsilon_{Ag}-\epsilon_{PMMA})^2}{12 \epsilon_\|^{(0)}} \left( \frac{t_{Ag}
t_{PMMA}}{t_{Ag} + t_{PMMA}}\frac{2\pi}{\lambda}\right)^2}.
\end{equation}
Here $\epsilon_\|^{(0)} = (t_{Ag} \epsilon_{Ag} + t_{PMMA} \epsilon_{PMMA})/(t_{Ag} + t_{PMMA})$ is the average of the metal and dielectric
permittivities $\epsilon_{Ag}$ and $\epsilon_{PMMA}$, respectively and $\lambda = 2\pi c/\omega$ is the vacuum wavelength. Note that
$\epsilon_\|^{(0)}$ is the permittivity predicted by the standard effective medium theory and that Eq.(\ref{permitt}) reduces to $\epsilon_\| =
\epsilon_\|^{(0)}$ in the limit of very small layers thicknesses $t_{Ag}$ and $t_{PMMA}$. From the denominator of Eq.(\ref{permitt}), it can be noted
that the real and imaginary parts of the metal and the polymer are correlated. This effect is only predicted by this improved description of the
effective medium, and is confirmed below by the experimental evidence. Using the Lorentz-Drude model for the silver permittivity $\epsilon_{Ag}$
reported in ref. [\onlinecite{RakicR}] and setting $\epsilon_{PMMA} = 1.95 \:$ \cite{NotaNo}, from Eq.(\ref{permitt}) we obtain
$\textrm{Re}(\epsilon_\|) = 0.0002$ at $\lambda = 595.4$ nm, for the chosen $t_{PMMA}$ and $t_{Ag}$.

From the data acquired through the measured T and R without optical pumping, we retrieved the real and imaginary parts of the dielectric
permittivities of the samples with $N=3,4,5$ bilayers, as shown in Fig. 2a and Fig. 2b, respectively.  In Fig. 2a we also plotted the theoretical
$\epsilon_\|$ of Eq.(\ref{permitt}) and $\epsilon_\|^{(0)}$. By inspecting the figure, it is evident that the permittivity of the sample with $N=3$
bilayers largely departs from the permittivities of the samples with $N=4$ and N=$5$ bilayers. The curves relative to these last two cases are almost
overlapped, agreeing better for longer wavelengths. It is then reasonable to conclude that the sample with $N=5$ bilayers can be regarded as a
homogeneous medium. Additionally, from the same figure it appears that the standard effective medium theory (dashed line) completely fails to
describe the dielectric response of the nano-laminates (as expected), whereas the improved theory of Eq.(\ref{permitt}) (solid line) provides an
accurate description. Figure 2c is a magnified copy of Fig. 2a in the spectral range $580$ nm $ < \lambda < 610$ nm, where $\textrm{Re}
(\epsilon_\|)$ changes sign, as predicted by the model and confirming the validity of our approximation.

\begin{figure}
\includegraphics[width=0.45\textwidth]{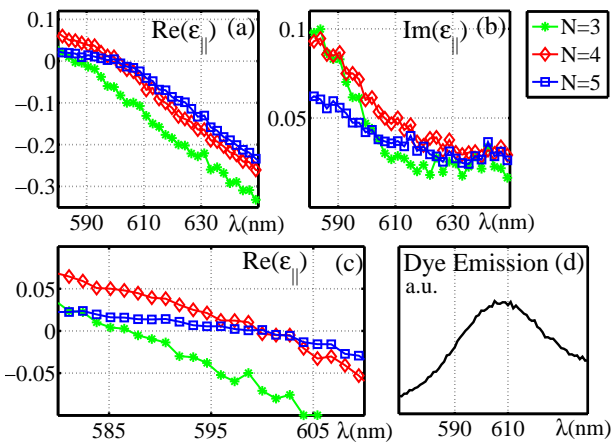}
\caption{(Color online) Retrieved dielectric permittivity $\epsilon_\|$ of optically pumped samples made up of $N=3,4,5$ metallo-dielectric bilayers,
as a function of the vacuum wavelength $\lambda$. (a) $\textrm{Re}(\epsilon_\|)$; (b) $\textrm{Im}(\epsilon_\|)$; (c) magnified copy of panel (a) in
the spectral range $580$ nm $ < \lambda < 610$ nm; (d) Dye molecules emission spectrum (arbitrary units).}
\end{figure}

The same retrieval procedure was repeated for the R and T data acquired in the presence of the optical pumping. The result is reported in Fig. 3. By
comparing Fig. 3b with Fig. 2b, we note that the imaginary part of the permittivity globally decreases over the spectral range where the dye
molecules are active (the emission spectrum of the dye is reported in Fig. 3d, in arbitrary units). Additionally, a close analysis of Fig. 3a shows
that the the optical gain has a limited effect on the real part of the effective permittivity of the samples. In particular, in the presence of
optical pumping, from Fig. 3c and for the sample with $N=5$, we note that the wavelength at which $\textrm{Re}(\epsilon_\|)$ vanishes is shifted to
$\lambda = 599$ nm. This shift confirms the validity of Eq.(\ref{permitt}), which correctly predicts the mixing of the real and imaginary part of the
permittivity.

This result manifestly shows that the minimization of the real part of the permittivity (ENZ condition) can be achieved together with the reduction
of its imaginary part, thus yielding an overall decrease of the absolute value of the sample effective permittivity. In Fig. 4 we plot the absolute
value of the permittivity of the sample with $N=5$ bilayers, both with and without optical pumping. The overall decrease of $| \epsilon_\| |$ is
particularly evident. In particular we obtain that, for the optically pumped sample, $| \epsilon_\| |$ attains its minimum value $0.04$ at $\lambda =
604$ nm, which corresponds to a $21.5$-percent decrease of the minimum $| \epsilon_\| | = 0.051$ (at $\lambda = 606$ nm) obtained without optical
pumping. Note that the minimum of $| \epsilon_\| |$ is not exactly attained at the wavelength where $\textrm{Re} (\epsilon_\|) \simeq 0$ (i.e. at
$\lambda = 599$ nm) and this is a consequence of the fact that the emission spectrum of the dye molecules has its maximum at $\lambda = 610$ nm (see
Fig.3d).

Finally, it should be noted that our setup allowed only for simple pump and probe experiments in cw, and in a not optimal absorption region of the
dye. However, we could register a linear dependence of T and R from the pump intensity (not shown), and we anticipate a much greater loss
compensation with an optimized optical pump.

In conclusion, we have experimentally investigated the homogenization of a nano-laminate with metal and dielectric gain constituents, with layers
thickness beyond the limit of validity of the standard effective medium theory. On the other hand, we have shown that the permittivity is correctly
described by an improved approach, which has allowed us to tailor the geometry to achieve the ENZ condition at a visible wavelength. In the presence
of optical pumping, we have shown that the imaginary part of the permittivity can be decreased, thus proving that the generalized ENZ where the
absolute value of the permittivity is very small can actually be observed. We predict that our results can open new ways to achieve efficient optical
steering based on the generalized ENZ condition $|\epsilon| \ll 1$.

We thank Salvatore Gambino for assistance with the choice of the dye and its characterization. ADF is supported by an EPSRC Career Acceleration
Fellowship (EP/I004602/1).

\begin{figure}
\includegraphics[width=0.45\textwidth]{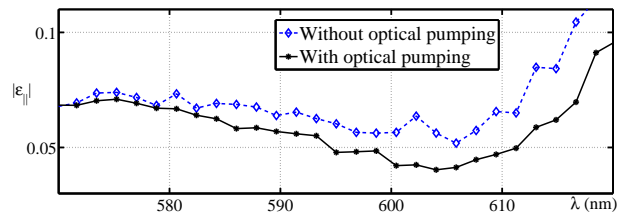}
\caption{(Color online) Absolute value of the permittivity of the sample with $N=5$ metal/polymer bilayers both with and without optical pumping.}
\end{figure}



\begin{thebibliography}{aa}
\bibitem{NotaNo} The value $\textrm{Re}(\epsilon_{PMMA}) = 1.95$ is slightly different from the standard value $\epsilon_{PMMA} = 2.22$
                 in order to consider the effect of the dye molecules. Besides, we neglected $\textrm{Im}(\epsilon_{PMMA})$ in the tailoring
                 of the structure.
\end{thebibliography}
\end{document}